\documentclass[a4paper,11pt]{article} 
\usepackage{cite}
\usepackage{setspace}
\usepackage{ae,aecompl,amsbsy,amssymb,eurosym}
\usepackage[english]{babel}
\usepackage{graphicx}
\usepackage{epsfig}
\usepackage{epstopdf}
\usepackage{amsmath}
\usepackage{amssymb}
\usepackage{subfigure}
\usepackage{ifthen}
\usepackage{alltt}
\usepackage{fancyhdr}
\usepackage{verbatim}
\usepackage{moreverb}
\usepackage{url}

\newcommand{\vt}[1]{\ensuremath{\mathbf{#1}}} 
\newcommand{\lt}[1]{\ensuremath{\mathrm{#1}}} 

\newcommand{\unitx}{\ensuremath{\vt{u}_x}}
\newcommand{\unity}{\ensuremath{\vt{u}_y}}

\newcommand{\unitr}{\ensuremath{\vt{u}_r}}
\newcommand{\unitt}{\ensuremath{\vt{u}_\theta}}
\newcommand{\unitp}{\ensuremath{\vt{u}_\phi}}

\newcommand{\adyad}{\ensuremath{\overline{\overline{\alpha}}}}
\newcommand{\tunitdyad}{\ensuremath{\overline{\overline{I}}}_\lt{t}}

\fancypagestyle{plain}{
  \fancyhead{} 
}

\begin{document}

\author{Antti~O.~Karilainen and Sergei~A.~Tretyakov\\
\\
Department of Radio Science and Engineering\\
/ SMARAD Center of Excellence,\\
Aalto University, P.~O.~Box 13000, FI-00076 Aalto, Finland.\\
Email: antti.karilainen@aalto.fi.}

\title{Circularly Polarized Receiving Antenna Incorporating Two Helices to Achieve Low Backscattering}

\maketitle

\begin{abstract}
We propose to use an antenna composed of two orthogonal helices as a low-scattering sensor. The vector effective length is derived for the antenna using the small dipole approximation for the helices. The antenna can transmit and receive circular polarization in all directions with the Huygens' pattern. We observe that the antenna geometry does not backscatter, regardless of the polarization, when the incidence direction is normal to the plane of the helices. Scattered fields, scattered axial ratio, and the scattering cross section are presented. We show that the zero-backscattering property holds also for the antenna when it is capable to receive all the available power with conjugate loading. The approximate analytical model is validated with full-wave simulations.
\end{abstract}



\section{Introduction}

When hiding a receiving antenna or an electromagnetic sensor from an observer the receiver must gather enough power and the antenna should not scatter too much power toward the observer. Antennas with only reactive loads can be made ``invisible'' in the dipole approximation by shaping the current distribution so that the induced dipole moment is zero. In \cite{Tretyakov2003a,Alu2010b}, this kind of an invisible object was realized by loading a wire dipole antenna with an inductive load. However, when an absorbing load (a receiver) is connected to the dipole, the antenna starts to scatter some power, as dictated by the forward scattering theorem. Due to the linear geometry of the antenna, it scatters power back to the incidence direction in the same intensity as in the forward direction. The ratio of the absorbed and scattered power can be made very high \cite{Alu2010b} at the expense of the received power. Another realization of this idea in the sensor design is the use of a plasmonic scattering cancellation technique \cite{Alu2009,Alu2010a}. These known solutions allow the control of the ratio of the received and scattered powers, but a compromise is always necessary: A nonzero received power implies nonzero scattered power in the direction toward the object under study. In this paper we propose to shape the scattering pattern of the receiving antenna and in addition to control the ratio between the received and scattered powers. This allows us to direct the unavoidable scattered power away from the observer, while allowing the antenna to receive all the available power from the incident wave. The method is conceptually very simple, and the proposed antenna uses only two metal wires of a specific shape.

In order to prevent backscattering, we need to have an antenna in which the induced dipole moments radiate away from the observer. Such a configuration is the combination of electric and magnetic dipole moments that replicates the magnitudes and phases of the incident wave's fields, also called the Huygens' source. A Huygens' source antenna for linear polarization (LP) is, for example, Green's antenna \cite{Green1966} which is a combination of a crossed dipole and loop antennas. Under conjugate loading, Green's antenna has the ability to receive power with the Huygens' pattern and scatter it to the opposite direction. Such minimum scattering antennas can inherently absorb the maximum amount of power that is equal to the total scattered power \cite{Kahn1965,Andersen2005,Kwon2009}. The antenna geometry we propose here uses two helices \cite{Tretyakov1996} aligned orthogonally, thus achieving in-plane isotropy for the antenna \cite{Alitalo2011}. In this study we consider this antenna theoretically and extend its possible use to sensor antennas by showing that the antenna can effectively absorb incident power, while directing the scattered power away from the incidence direction (as proposed in \cite{Karilainen2011}).


\section{Receiving Antenna}
\label{sec:receiving_antenna}

\begin{figure}[!t]
  \centering
  \includegraphics[width=80mm]{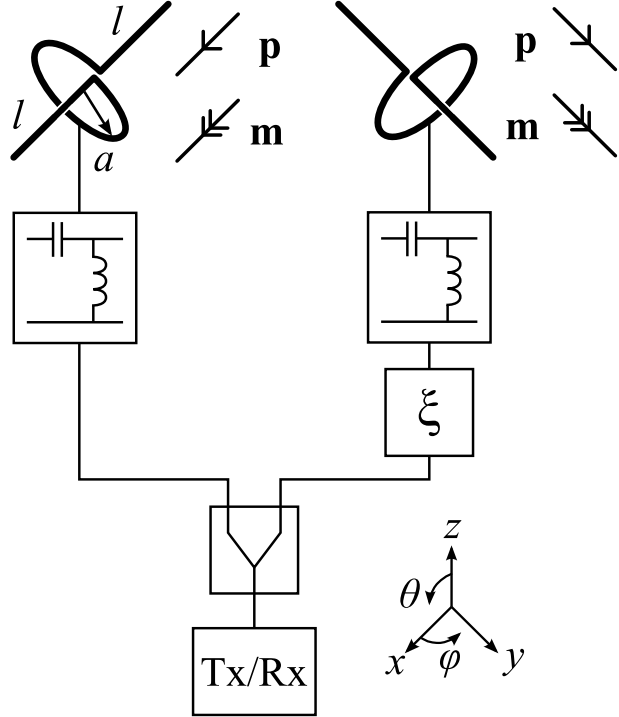}
  \caption{Electric and magnetic sources $\vt{p}$ and $\vt{m}$, respectively, corresponding to the antenna formed by two helices (see Fig.~\ref{fig:simulation_model} for the actual placement). The helices are connected to the receiver (Tx/Rx) via a power divider, $\xi$ phase shift in the second helix, and optional matching circuits.}
  \label{fig:elem_sources}
\end{figure}

The antenna geometry has been studied in the transmitting mode in \cite{Alitalo2011} according to Fig.~\ref{fig:elem_sources} where the helices are drawn, for clarity, separately (see Fig.~\ref{fig:simulation_model} for the actual placement). We model two right-handed (RH) helices with electrically short dipoles with the length $2l$ (triangular current distribution) and small conducting loops with the radius $a$, loop area $S=\pi a^2$ (uniform current distribution). The sources are aligned so that the electric and magnetic dipoles point in both $x$ and $y$ directions. Both of the helices are assumed to be located in the origin and the wire and loop parts to be balanced with respect to the scattered power using $Sk = l$, where $k = \omega\sqrt{\mu_0 \epsilon_0}$.

We can study the impedance matching, polarization matching, and the absorbed power in the load from an incident electric field $\vt{E}_\lt{inc}$ easily with the equation \cite{antenniteoria}:
\begin{equation}
    \centering
    P_\lt{abs} = \frac{4 R_\lt{a} R_\lt{L}}{|Z_\lt{a} + Z_\lt{L}|^2} \frac{|\vt{h} \cdot \vt{E}_\lt{inc}|^2}{|\vt{h}|^2 |\vt{E}_\lt{inc}|^2} \frac{\eta |\vt{h}|^2}{4R_\lt{a}} \frac{|\vt{E}_\lt{inc}|^2}{2\eta} ,
    \label{eq:P_L}
\end{equation}
where $R_\lt{a}$ is the real part of the  antenna impedance $Z_\lt{a}$, $R_\lt{L}$ is the real part of the complex load impedance $Z_\lt{L}$, \vt{h} is the vector effective length, $\vt{E}_\lt{inc}$ is the incident electric field, and $\eta = \sqrt{\mu_0/\epsilon_0}$.

For short dipoles and small loops $\vt{h}$ are known (e.g., \cite{Balanis2005}), and we can write the receiving pattern $\vt{F}_\lt{rx}$ using $\vt{h}$ for the antenna as a function of the phasing between the helices $\xi$ as:
{\setlength\arraycolsep{0pt}
\begin{eqnarray}
    |\vt{h}|^2 = l^2 |\vt{F}_\lt{rx}|^2 = 2 l^2 \big[ & & \cos^2\theta + \cos\xi \sin 2\phi (\cos^2\theta - 1) \nonumber \\
     & &  +  2\sin\xi \cos\theta + 1
    \big ].
    \label{eq:Frx}
\end{eqnarray}
If there is no phase shift ($\xi=0$), $\vt{F}_\lt{rx}$ is the pattern of a small dipole pointing at $\theta = 90^\circ$, $\phi = 45^\circ$ with the maximum directivity $D_\lt{max} = 1.5$ (1.76~dB). A Huygens' source pattern with $D_\lt{max} = 3$ (4.77~dB) in the $+z$ direction and a null in the $-z$ direction is achieved with a $+90$-degree phase shift. The Huygens' pattern flips over at $-z$ if the phase difference is $-90$ degrees.

Let the incident field be arbitrarily polarized with an elliptic component according to $\vt{E}_\lt{inc} = E_\lt{inc}(-\vt{u}_\theta \pm jAR_\lt{inc}\vt{u}_\phi)\exp(jkr)$, where $E_\lt{inc}$ is the magnitude and $AR_\lt{inc}$ is the inverted axial
ratio ($0 \leq AR_\lt{inc} \leq 1$, $AR_\lt{inc} = 0$ for LP, $AR_\lt{inc} = 1$ for CP). The top sign ``$\pm$'' is for left-handed (LH) and the bottom sign for RH polarization. We can calculate the polarization efficiency for the Huygens' reception pattern as
\begin{equation}
    p_\lt{e} = \frac{ |\vt{h} \cdot \vt{E}_\lt{inc}|^2 }{ |\vt{h}|^2 |\vt{E}_\lt{inc}|^2 } = \frac{1}{2} \frac{(AR_\lt{inc} \mp 1)^2}{1 + AR_\lt{inc}^2},
    \label{eq:polmatch}
\end{equation}
which can be seen to be independent of the incident angle. This is an expected results, as the antenna of RH helices also transmits right-hand circular polarization (RHCP) in all directions in the ideal case \cite{Alitalo2011}.


\section{Scattering From the Antenna}
\label{sec:scattering}

\subsection{Scattering From the Lossless Structure}

Scattering from the lossless antenna structure is calculated via the induced electric and magnetic dipole moments $\vt{p}$ and $\vt{m}$, respectively, as follows:
{\setlength\arraycolsep{2pt}
\begin{eqnarray}
    \vt{p} & = & \adyad_\lt{ee} \cdot \vt{E}_\lt{inc} + \adyad_\lt{em} \cdot \vt{H}_\lt{inc},
    \label{eq:psimple} \\
    \vt{m} & = & \adyad_\lt{me} \cdot \vt{E}_\lt{inc} + \adyad_\lt{mm} \cdot \vt{H}_\lt{inc},
    \label{eq:msimple}
\end{eqnarray}
where $\vt{E}_\lt{inc}$ and $\vt{H}_\lt{inc}$ are the exciting fields, and $\adyad$ are the polarizabilities written using the dyadic notation \cite{Serdyukov2001}. Scattering from a single helix has been studied in \cite{Tretyakov1996}, where complete scattering components were presented. Here we study, for the sake of simplicity, the scattering from the antenna structure of two helices so that we neglect the electric polarizability of the loop. We are left with only four polarizability components from \cite{Tretyakov1996}. For two identical helices the polarizabilities have the form
{\setlength\arraycolsep{2pt}
\begin{eqnarray}
	\adyad_\lt{ee} & = & \alpha_\lt{ee} \tunitdyad, \quad \adyad_\lt{em} = \alpha_\lt{em} \tunitdyad, \nonumber \\
	\quad \adyad_\lt{me} & = & -\adyad_\lt{em}, \quad \adyad_\lt{mm} = \alpha_\lt{mm} \tunitdyad,
	\label{eq:pols}
\end{eqnarray}
where $\tunitdyad = \unitx\unitx + \unity\unity$.
The polarizabilities are, using $Sk = l$, the following~\cite{Tretyakov1996}:
{\setlength\arraycolsep{2pt}
\begin{eqnarray}
	\alpha_\lt{ee} & = & \frac{l^2}{j\omega}\frac{1}{Z_\lt{a}}, \label{eq:a_ee} \\	
	\alpha_\lt{em} & = & \pm \eta \frac{l^2}{\omega}\frac{1}{Z_\lt{a}} = \pm j\eta \alpha_\lt{ee} \label{eq:a_em}, \\
	\alpha_\lt{me} & = & -\alpha_\lt{em} = \mp j\eta \alpha_\lt{ee}, \label{eq:a_me} \\
	\alpha_\lt{mm} & = & \eta^2 \frac{l^2}{j\omega}\frac{1}{Z_\lt{a}} = \eta^2 \alpha_\lt{ee}, \label{eq:a_mm}
\end{eqnarray}
where $Z_\lt{a}$ is the impedance of the helix. In (\ref{eq:a_em}) the ``$\pm$'' sign determines the handedness, bottom sign for RH helices. The scattered far field is calculated from (\ref{eq:psimple}) and (\ref{eq:msimple}) according to (e.g.~\cite{Lindell2009})
\begin{equation}
	\vt{E}_\lt{sca} = -\omega^2 \mu_0 \frac{e^{-jkr}}{4 \pi r} \bigg( \unitr \times (\unitr \times \vt{p}) + \unitr \times \frac{\vt{m}}{\eta}  \bigg).
	\label{eq:Esca}
\end{equation}

Let us assume an LP incident plane wave arriving at the structure from the direction ($\theta_\lt{inc}$, $\phi_\lt{inc}$), so that it is TM polarized with respect to the plane of the helices. The complex scattering pattern can be calculated using (\ref{eq:Esca}) as
\begin{equation}
    \vt{F}_\lt{sca}^\lt{TM} = \unitt\Theta_\lt{sca}^\lt{TM} + \unitp\Phi_\lt{sca}^\lt{TM},
    \label{eq:FscaTM}
\end{equation}
where $\Theta_\lt{sca}^\lt{TM}$ and $\Phi_\lt{sca}^\lt{TM}$ are the $\theta$ and $\phi$ components, respectively. The magnitude pattern is solved as:
\begin{equation}
    |\vt{F}_\lt{sca}^\lt{TM}| = \sqrt{|\Theta_\lt{sca}^\lt{TM}|^2 + |\Phi_\lt{sca}^\lt{TM}|^2},
    \label{eq:absFscaTM}
\end{equation}
but let us first examine the intermediate result
\begin{eqnarray}
    \lefteqn{ |\Theta_\lt{sca}^\lt{TM}|^2 + |\Phi_\lt{sca}^\lt{TM}|^2 = {} } \nonumber\\
        &   & 2\cos^2\phi(\cos^2\theta + 1) (\cos^2\phi_\lt{inc}\cos^2\theta_\lt{inc} + \sin^2\phi_\lt{inc}) \nonumber \\
        & + & 2\sin^2\phi(\cos^2\theta + 1) (\sin^2\phi_\lt{inc}\cos^2\theta_\lt{inc} + \cos^2\phi_\lt{inc}) \nonumber \\
        & - & 4\sin\theta\cos\phi(\cos^2\theta - 1) \sin\phi_\lt{inc}\cos\phi_\lt{inc} (\cos^2\theta_\lt{inc} - 1) \nonumber \\
        & - & 4\cos\theta \cos\theta_\lt{inc}.
    \label{eq:sca_components}
\end{eqnarray}
For the normal incidence from the $\pm z$ direction, we set $\cos\theta_\lt{inc} = \pm1$, and see from (\ref{eq:sca_components}) that the change in the incident direction affects only the last term. For a plane wave from the $+z$ direction $|\vt{F}_\lt{sca}^\lt{TM}| \propto (\cos\theta - 1)$, and for a plane wave from the $-z$ direction $|\vt{F}_\lt{sca}^\lt{TM}| \propto (\cos\theta + 1)$. In other words, for the normal incidence the balanced structure always forward scatters as an ideal Huygens' source, without any backscattering. The fundamental reason between zero-backscattering is the self-duality according to (\ref{eq:a_mm}) in addition to the in-plane isotropy of the structure~\cite{Lindell2009}.

For oblique incidence the pattern starts to resemble the pattern of an elementary dipole. For example, if we set $\theta_\lt{inc} = \pi/2$ and $\phi_\lt{inc} = 0$, i.e.\ incidence from the $+x$ direction, the pattern is that of a $y$-directed elementary dipole, and the forward and backscattering magnitudes are equal. Moreover, the scattering behavior is isotropic in the $xy$ plane: the antenna structure always scatters the same pattern for all $\phi_\lt{inc}$, when observed from the incident direction. Scattering directivity is plotted in Fig.~\ref{fig:dir_sca_db} for small incident angles in decibel scale.

\begin{figure}[!t]
  \centering
  \includegraphics[width=120mm]{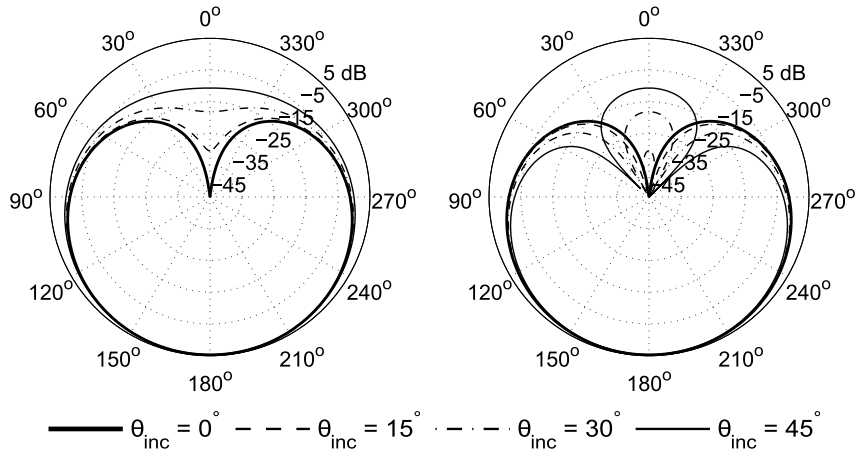}
  \caption{Scattering directivity (decibel scale) for the structure when the incident TM-polarized plane wave arrives at any $\phi_\lt{inc}$ and various $\theta_\lt{inc}$. Parallel (left) and perpendicular (right) $\theta$ plane relative to $\phi_\lt{inc}$.}
  \label{fig:dir_sca_db}
\end{figure}

The axial ratio of the scattered field can be of interest in some applications, and it is easily obtainable from the complex scattering pattern. We use the $\vt{p}_\lt{h}$~vector~\cite{Lindell1996}
\begin{equation}
    \vt{p}_\lt{h} = \frac{\vt{F}_\lt{sca}^\lt{TM} \times \big( \vt{F}_\lt{sca}^\lt{TM} \big)^* }{ j \vt{F}_\lt{sca}^\lt{TM} \cdot \big( \vt{F}_\lt{sca}^\lt{TM} \big)^* },
    \label{eq:pvector}
\end{equation}
where ${}^*$ stands for complex conjugate. If $|\vt{p}_\lt{h}| = 1$ in (\ref{eq:pvector}), we have CP; for $0 < |\vt{p}_\lt{h}| < 1$ we have elliptic polarization; and for $|\vt{p}_\lt{h}| = 0$ we have LP. By substituting  (\ref{eq:FscaTM}) to (\ref{eq:pvector}), we get $\vt{p}_\lt{h} = \vt{u}_\lt{r}$ and $AR=1$ independent of the incident and scattering directions. The scattered field is RHCP as $\vt{p}_\lt{h}$ is directed along $\vt{u}_\lt{r}$, and if the helices are LH, we get left-hand circular polarization (LHCP) scattered wave as $\vt{p}_\lt{h} = -\vt{u}_\lt{r}$.

The calculations above were performed for a TM-polarized incident plane wave, but exactly the same results are valid for TE and CP polarizations as well, as a CP wave is a combination of TM and TE waves. The only exception happens naturally when LHCP (RHCP) incident wave arrives at the antenna formed by RH (LH) helices, in which case no response is seen. It can be however noticed that if we let the polarization of an elliptic LH incident wave approach LHCP, we see that the scattered field handedness follows the same trend as for the linear polarization: the handedness of the antenna structure determines the handedness of the scattered field.


\subsection{Scattering From the Loaded Structure}
\label{sec:scattering_loaded}

In \cite{Karilainen2011} it was suggested that the structure can be also used as a receiving antenna. In reception the analytical model for scattering becomes more complicated than (\ref{eq:Esca}) due to the additional current induced by the load. In the general case the scattered field for an antenna with a complex load $Z_\lt{L}$ can be written as \cite{Balanis2005}
\begin{equation}
    \vt{E}_\lt{sca}(Z_\lt{L}) = \vt{E}_\lt{sca} - \frac{I_\lt{sca}}{I_\lt{tx}} \frac{Z_\lt{L}}{Z_\lt{L} + Z_\lt{a}} \vt{E}_\lt{tx},
    \label{eq:Esca_loaded}
\end{equation}
where $\vt{E}_\lt{sca}$ is the scattered electric field without a load, (\ref{eq:Esca}); $I_\lt{sca}$ is the current at the position of the load induced by the incoming field; $I_\lt{tx}$ is the current driven by the load; and $\vt{E}_\lt{tx}$ is the electric field transmitted by the structure with the current $I_\lt{tx}$. The current that flows in the wire and loop can be calculated from the induced current in the wire \cite{Tretyakov1996}: $I_\lt{sca} = E_\lt{inc} l/Z_\lt{a}$ (due to the balance condition, the same current is induced in the loop). By inserting the currents $I_\lt{sca}$ and $\vt{E}_\lt{tx}/I_\lt{tx}$ in (\ref{eq:Esca_loaded}), we find:
\begin{equation}
    \vt{E}_\lt{sca}(Z_\lt{L}) = \frac{e^{-jkr}}{4\pi r}  j\omega\mu l \frac{ E_\lt{inc} l}{Z_\lt{a}} \bigg[ \vt{F}_\lt{sca} - \frac{Z_\lt{L}}{Z_\lt{L} + Z_\lt{a}} \vt{F}_\lt{tx} \bigg], \label{eq:Esca_loaded2}
\end{equation}
where the scattering pattern $\vt{F}_\lt{sca}$ includes the components from the incident TM wave (\ref{eq:FscaTM}) and a similar TE wave, and the pattern in the ``transmission mode'' is $\vt{F}_\lt{tx}$ (not related to the reception pattern $\vt{F}_\lt{rx}$).

Here and in the following we assume that the antenna scatters as a combination of two electric and two magnetic dipoles, neglecting higher-order current modes due to scattering from antenna loads. Because the currents contributing to the lossless and loaded structure have different distributions along the wire and the load impedances of the two helices can be in general different, the field patterns $\vt{F}_\lt{tx}$ and $\vt{F}_\lt{sca}$ are different functions of the observation angles. However, within the adopted electric and magnetic dipole-moment approximation, neglecting the possible higher-order modes, we can approximately set $\vt{F}_\lt{tx} \approx \vt{F}_\lt{sca}$ for the same loading in both helices. This means that the scattering pattern is again the Huygens' pattern, and the received power is dictated by the usual matching factor for power in (\ref{eq:P_L}). If the load $Z_\lt{L}$ perturbs significantly the current distribution along the wire, the resulting multipole source has to be found by using more sophisticated methods, such as the method of moments, in order to calculate $\vt{F}_\lt{tx}$ accurately.

The normalized scattering cross section can be calculated from (\ref{eq:Esca_loaded2}), and is as follows:
\begin{eqnarray}
    \sigma/\lambda^2 & = & \frac{1}{\lambda^2} 4\pi r^2 \frac{|\vt{E}_\lt{sca}(Z_\lt{L})|^2}{|\vt{E}_\lt{inc}|^2} \nonumber \\
                 & = & \frac{1}{16 \pi^2} k^4 l^4 \frac{\eta^2}{|Z_\lt{a}|^2} \bigg|  \vt{F}_\lt{sca} - \frac{Z_\lt{L}}{Z_\lt{L} + Z_\lt{a}} \vt{F}_\lt{tx} \bigg|^2.
    \label{eq:rcs}
\end{eqnarray}


\subsection{Absorbed and Scattered Power}

It is known that the ratio between the  absorbed and scattered powers is related to the reception directivity and scattering directivity \cite{Andersen2005} as well as to the load impedance \cite{Alu2010b}. Here we determine this ratio using the model presented in this paper. The absorbed power can be calculated from (\ref{eq:P_L}):
\begin{eqnarray}
    P_\lt{abs} & = & \frac{R_\lt{L}}{|Z_\lt{a} + Z_\lt{L}|^2} \frac{p_\lt{e}}{2} |\vt{h}|^2 |\vt{E}_\lt{inc}|^2 \nonumber \\
    & = & \frac{R_\lt{L}}{|Z_\lt{a} + Z_\lt{L}|^2} \frac{|\vt{F}_\lt{rx}|^2}{4}  E_\lt{inc}^2 l^2 (AR_\lt{inc} \mp 1)^2,
    \label{eq:Pabs}
\end{eqnarray}
where $\vt{h}$ and $\vt{E_\lt{inc}}$ are as before. The scattered power is obtained by integrating the Poynting vector of the scattered far field (\ref{eq:Esca_loaded2}) over a closed surface $S$:
\begin{eqnarray}
    P_\lt{sca} & = & \oint_S \frac{|\vt{E}_\lt{sca}(Z_\lt{L})|^2}{2\eta} \lt{d} S \nonumber \\
    & = & \frac{1}{32\pi} \eta k^2 l^2 \frac{E_\lt{inc}^2 l^2 (AR_\lt{inc} \mp 1)^2}{|Z_\lt{a} + Z_\lt{L}|^2} \oint_S |\vt{F}_\lt{sca}|^2 \lt{d} \Omega \nonumber \\
    & = & \frac{2}{3\pi} \eta k^2 l^2 \frac{E_\lt{inc}^2 l^2 (AR_\lt{inc} \mp 1)^2}{|Z_\lt{a} + Z_\lt{L}|^2},
    \label{eq:Psca}
\end{eqnarray}
where $\vt{F}_\lt{tx} = \vt{F}_\lt{sca}$ and normal incidence were assumed. The scattered power (\ref{eq:Psca}) has also the term $(AR_\lt{inc} \mp 1)^2$ which takes into account the polarization matching between the incident wave and the handedness of the antenna. The ratio between $P_\lt{abs}$ and $P_\lt{sca}$ is then
\begin{equation}
    \frac{P_\lt{abs}}{P_\lt{sca}} = 8 \pi^2 \frac{1}{k^2 l^2} \frac{R_\lt{L}}{\eta} \frac{ |\vt{F}_\lt{rx}|^2 }{ \oint_S |\vt{F}_\lt{sca}|^2 \lt{d} \Omega } = \frac{3 \pi}{8} \frac{1}{k^2 l^2} \frac{R_\lt{L}}{\eta} |\vt{F}_\lt{rx}|^2.
    \label{eq:Pratio}
\end{equation}
When deriving (\ref{eq:Pratio}) we have only assumed that the scattered pattern from the antenna is approximately the same as that of the lossless structure (see Section~\ref{sec:scattering_loaded}). In other words, if we stay in the same mode with no backscattering, $P_\lt{abs}/P_\lt{sca}$ can be theoretically arbitrary, depending only on the real part $R_\lt{L}$ of the load impedance $Z_\lt{L}$ and the reception pattern.

Reflecting the results of \cite{Andersen2005}, we  should see that the ratio (\ref{eq:Pratio}) is related to the directivities of $\vt{F}_\lt{rx}$ and $\vt{F}_\lt{tx} = \vt{F}_\lt{sca}$ when the antenna is terminated with a conjugate load. Assuming that the helices are at resonance, we need to solve the radiation resistance of the antenna to implement the conjugate matching. The radiation resistance of a lossless short dipole with a triangular current distribution $R_\lt{w}$ and a small loop $R_\lt{l}$ are known (e.g.~\cite{Balanis2005}), and due to the balance condition $Sk = l$, it follows that $R_\lt{w} = R_\lt{l}$, and the total radiation resistance for the antenna is
\begin{equation}
    R_\lt{a} = R_\lt{w} + R_\lt{l} = \frac{\eta}{3\pi} k^2 l^2.
    \label{eq:R_a}
\end{equation}
Finally, we can insert $R_\lt{L} = R_\lt{a}$ to (\ref{eq:Pratio}), and get simply
\begin{equation}
    \frac{P_\lt{abs}}{P_\lt{sca}} = \frac{|\vt{F}_\lt{rx}|^2}{8}.
    \label{eq:Pratio_conjload}
\end{equation}
Now, assuming the Huygens' reception pattern with 90 degrees phase shift between the helices in (\ref{eq:Frx}) and normal incidence, the ratio is $P_\lt{abs}/P_\lt{sca} = 1$. This is an expected results, as the reception and scattering patterns are identical, but directed in the opposite directions \cite{Andersen2005}. If no phase shift is introduced between the helices, we get $P_\lt{abs}/P_\lt{sca} = 1/2$, as half of the available power is dissipated in the power divider using the scheme of Fig.~\ref{fig:elem_sources}.

In the conjugate-loading case the maximum absorbed power for the normal incidence is found by substituting (\ref{eq:R_a}) and (\ref{eq:Frx}) with the Huygens' pattern in (\ref{eq:Pabs}) as:
\begin{equation}
    P_\lt{abs}^\lt{max} = \frac{|\vt{h}|^2 |\vt{E}_\lt{inc}|^2}{16 R_\lt{a}} = \frac{|\vt{E}_\lt{inc}|^2}{2 \eta} \frac{3 \pi |\vt{h}|^2}{8 k^2l^2} = \frac{|\vt{E}_\lt{inc}|^2}{2 \eta} \frac{3 \lambda^2}{4 \pi},
    \label{eq:Pabs_conjload}
\end{equation}
which is the limit for maximum received power for small antennas derived in \cite{Kwon2009}.


\section{Numerical Validation}

To validate the developed model, the scattered fields from the antenna were simulated using the integral equation method in ANSYS HFSS 13 full-wave simulator. The antenna was modeled using PEC wire with a square cross-section (the edge length is $w=0.17$~mm). The arm length is $l$ and the gap between the arms is kept constant $g=0.8$~mm. The loop is modeled as a one-turn helix with a pitch of 1.0~mm, making it possible to align both of the helices symmetrically in the origin. The helical loops with the radius $a$ cross one another at the bottom of the loop where PEC strips (length 0.5~mm) were used to prevent the two helices from touching each other while maintaining the loop size. The simulation model is illustrated in Fig.~\ref{fig:simulation_model}. The total length of the wire $L_\lt{tot}$ of each helix has the electrical  length close to half the wavelength $\lambda_0/2$ at $f_0=1.3$~GHz, which was used as the starting point in the simulations. Helix dimensions were found for minimum backscattering for the normal incidence at $f_0$ by tuning $a$ and $l$, or in other words, by optimizing the magnitude of radiation from the induced electric and magnetic dipole moments to be equal in the far zone.

\begin{figure}[!t]
  \centering
  \includegraphics[width=80mm]{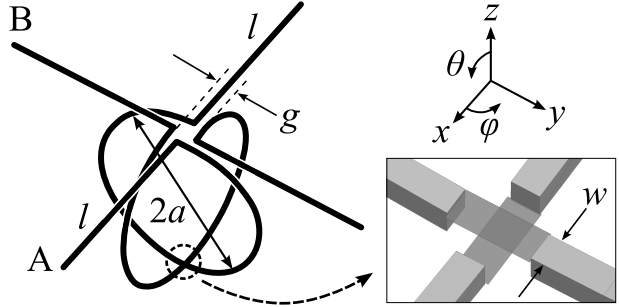}
  \caption{Simulation model for the antenna. The loops are modeled as helices with a short pitch, thus enabling the helices to lie orthogonally relative to each other. Loads at the bottom of the loops are modeled as impedance sheets.}
  \label{fig:simulation_model}
\end{figure}

\subsection{Simulations for the Lossless Structure}
\label{sec:validation_lossless}

The found dimensions for the lossless structure are $l=21.01$~mm and $a=10.50$~mm. The scattering directivities are presented in Fig.~\ref{fig:hfss_dir_sca_db_tm} for a linear TM-polarized plane wave. The scattered lobe is somewhat tilted toward the incidence direction at the oblique $\theta_\lt{inc}$ angles. Also, the patterns show some asymmetry in the orthogonal scattering plane, because in the simulations the pattern is twisted a few degrees around the $z$ axis. The scattering levels in the $+z$ direction at oblique angles are similar to the theoretical ones in Fig.~\ref{fig:dir_sca_db}. We do not observe the expected perfect RHCP in the forward scattering direction, but see only the inverted axial ratio of 0.7 (simulation results not shown here). However, this applies only to LP incidence, and in \cite{Karilainen2011c} the scattered polarization in the forward direction was CP when the incident field was CP as well.

\begin{figure}[!t]
  \centering
  \includegraphics[width=120mm]{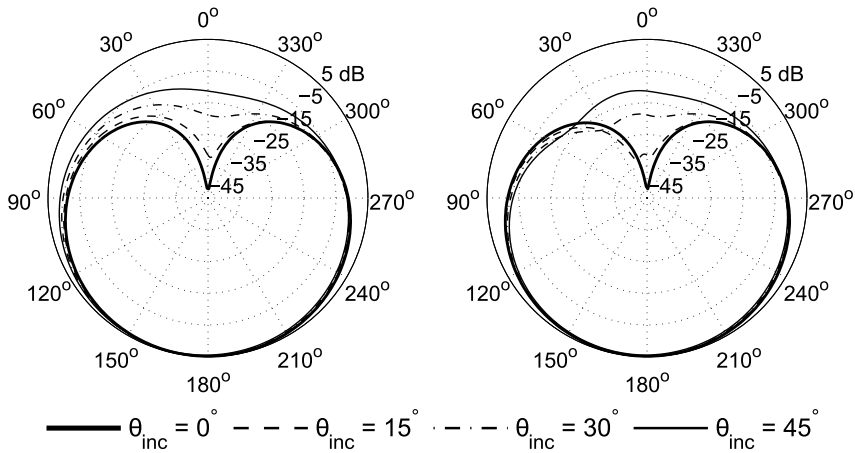}
  \caption{Simulated scattering directivity (decibel scale) for the lossless structure when the incident TM-polarized plane wave arrives at $\phi_\lt{inc}=0^\circ$ and various $\theta_\lt{inc}$. Parallel (left) and perpendicular (right) $\theta$ plane relative to $\phi_\lt{inc}$.}
  \label{fig:hfss_dir_sca_db_tm}
\end{figure}


\subsection{Simulations for the Loaded Structure}
\label{sec:sim_conjload}

The receiving antenna fulfills the conjugate-matching criterion to receive maximum available power according to (\ref{eq:P_L}), if the load connected to the antenna is $Z_\lt{L} = Z_\lt{a}^*$ when neglecting the material losses. Reactive impedance sheets of $C=2.15$~pF were added to the simulation model to tune the helices to resonance. As reactive loading changes the current distribution along the wires, we find new values for minuscule backscattering as $l=19.66$ and $a=10.93$~mm. The patterns are similar as before, but the polarization is now close to CP in the forward direction, as seen in Fig.~\ref{fig:hfss_ar_sca_lossless_tuned} that shows the inverted axial ratio. Finally, resistive sheets of $R_\lt{L}=R_\lt{a}=2.88$~$\Omega$ were added to realize the conjugate load. The scattering directivities with reactive and conjugate load as well as with TE incidence polarization are similar to Fig.~\ref{fig:hfss_dir_sca_db_tm}, and not repeated here.

\begin{figure}[!t]
  \centering
  \includegraphics[width=120mm]{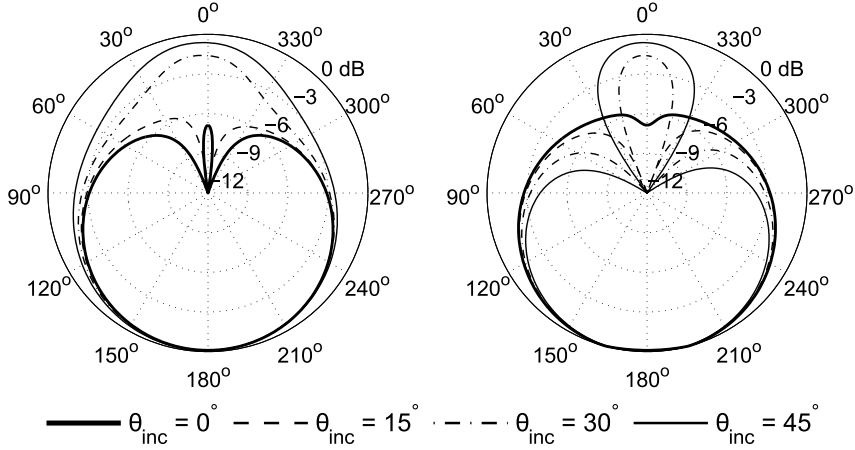}
  \caption{Simulated scattered inverted axial ratio for the tuned and lossless structure when the incident TM-polarized plane wave at normal incidence is parallel (left) and perpendicular (right) $\theta$ plane relative to $\phi_\lt{inc}$.}
  \label{fig:hfss_ar_sca_lossless_tuned}
\end{figure}

The simulated values of $\sigma/\lambda^2$ for the normal incidence are shown in Fig.~\ref{fig:rcs_comparison} for the lossless structure, for the lossless structure at resonance, and for the conjugate-loaded antenna. The analytical results from (\ref{eq:rcs}) are shown for comparison. In the analytical results we have used the antenna impedance $Z_\lt{a}$ and the loop radius $a$ from the simulations with the balance condition $Sk=l$. The results are in agreement, and in principle the geometry can be optimized to produce as low levels of backscattering as desired, although numerical accuracy plays an increasing role with low field magnitudes.

\begin{figure}[!t]
  \centering
  \includegraphics[width=120mm]{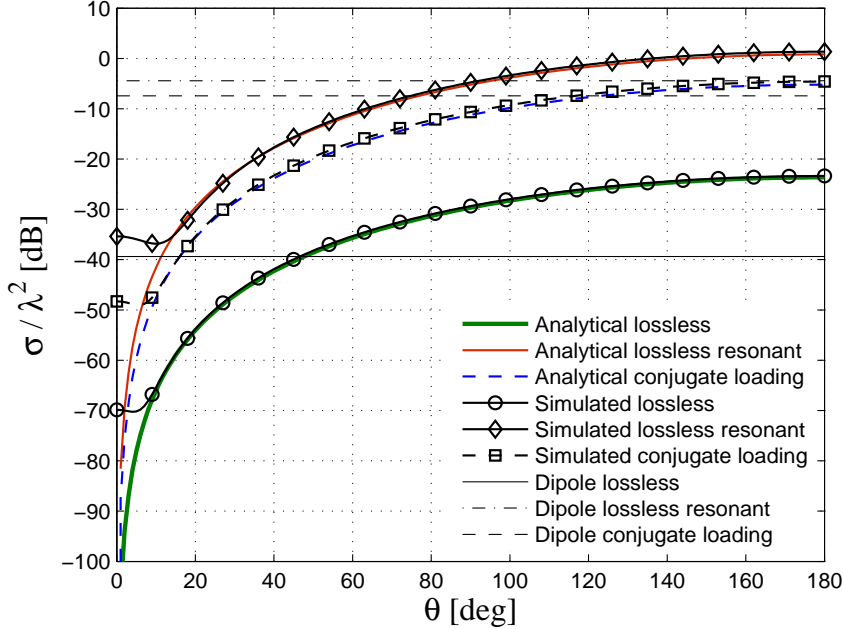}
  \caption{Analytical and simulated $\sigma/\lambda^2$ for the lossless structure and the conjugate-loaded antenna with reference-dipole limits.}
  \label{fig:rcs_comparison}
\end{figure}

Additionally, Fig.~\ref{fig:rcs_comparison} shows the simulated reference values for short dipoles, that have the same electric length as the wire parts of the helices. The dipoles were simulated with the total lengths of $2l+g=\lambda/5.39$ for the lossless case and $\lambda/5.75$ for the resonant cases. To summarize, the antenna structure backscatters at least 30~dB less in the lossless case, and at least 40~dB less in the conjugate-loaded case when compared to a small dipole with similar electrical dimensions at normal incidence.

How wide is the frequency range where this antenna has low backscattering levels? To answer this question, we simulate the backscattering cross section over a frequency range close to the designed $f_0$, shown in Fig.~\ref{fig:rcs_element_dipole_mono_comparison} for the three different cases. The lossless structure exhibits the widest variations of the backscattering levels. At its resonant frequency ($0.946 f_0$) the necessary balance between the induced electric and magnetic dipole moments is not satisfied, and scattering is strong. In contrast, backscattering at $f_0$ is very small because the Huygens' condition is satisfied and in addition the induced currents are weak as this is not the resonant frequency of the helices. When the structure is tuned to resonance at $f_0$, the resonance curve transforms into a gentle slope with some variation in the vicinity of $f_0$, because the effects of these two factors essentially compensate each other. In the optimal case, the backscattering would go to zero at $f_0$ with the resonant structures, but in Fig.~\ref{fig:rcs_element_dipole_mono_comparison} the simulation and geometry unidealities distort the curve. Simulated values for the individual reference dipoles are presented for comparison, and it is seen that the antenna scatters as little as the non-resonant dipole without any loading. Although the low scattering levels appear over quite a wide frequency range, the accepted power is dictated by the conjugate matching and is naturally at its maximum only at $f_0$ in our case.

\begin{figure}[!t]
  \centering
  \includegraphics[width=120mm]{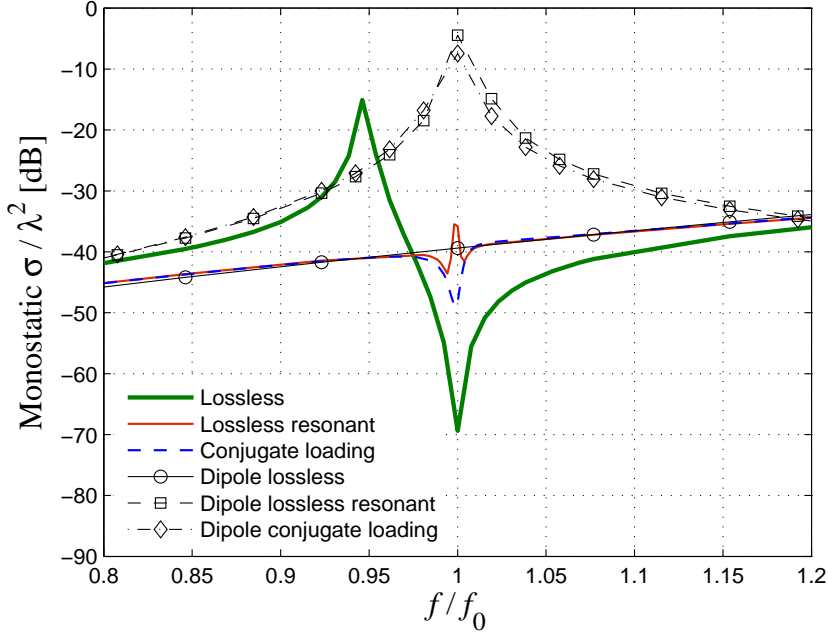}
  \caption{Simulated monostatic scattering cross section for normal incidence for the lossless, lossless at resonance, and conjugate-loaded structure with the reference dipoles.}
  \label{fig:rcs_element_dipole_mono_comparison}
\end{figure}


\section{Discussion}

In this paper we have extended the analysis of the antenna structure of \cite{Alitalo2011} from transmission to reception and scattering. The results in terms of the scattered magnitude and scattering cross section are well in agreement with simulations regardless of the approximations in the model. The scattered polarization from an LP incident wave was circular in the forward scattering direction only when the helices were tuned to resonance, a similar effect as predicted in \cite{Wheeler1958,Thal2006} for a spherical helix geometry. The approximate analytical model of this paper is not consistent in terms of conservation of power when compared to the wire dipole in \cite{Alu2010b}, but because the structure is relatively small, the deviations from the ideal case are small.

The absorbed and scattered powers were calculated, and the ratio was seen to be in agreement with the theory of absorption efficiency \cite{Andersen2005,Kwon2009} in case of conjugate loading. The antenna presented here appears to be an optimal small receiving antenna---for both linear and circular polarizations. Additionally, it was seen that the ratio between the absorbed and scattered power in (\ref{eq:Pratio}) is proportional to the load resistance, offering some control over the ratio as long as the dipole approximation holds.

This antenna geometry may find other uses besides antenna applications. For example, an array of such geometry that forms the dipole moments (or, equivalent currents) directing the scattered field away from the incident direction can be possibly used as a matching layer in applications where power is collected or absorbed. The shape of the helix is flexible, and it can be designed to be, for example, semi-planar as in \cite{Karilainen2011} or a genuine helix. Also, when operating with the balance condition $Sk = l$ at the fundamental resonance, the ratio between the wire length and the loop radius is constant regardless of the resonant frequency, i.e., the helix shape stays the same and can be scaled to all microwave frequencies.


\section{Conclusion}

A lossless antenna structure formed by two helices possesses an ideal combination of induced dipole moments that direct the scattered field away from the incident direction. It was seen that this property holds when the antenna received power. The absorbed and scattered powers were solved, and it was seen that the antenna is an optimal small receiving antenna for linear and circular polarization.

\section*{Acknowledgment}

A.~O.~Karilainen would like to thank T.~Niemi and P.~Alitalo for useful discussions. A.~O.~Karilainen and S.~A.~Tretyakov acknowledge the support of the Academy of Finland and Nokia Corporation through the Center-of-Excellence program. The work of A.~O.~Karilainen has been supported by the Graduate School, Faculty of Electronics, Communications, and Automation; HPY Research Foundation; and Elektroniikkainsin\"o\"orien s\"a\"ati\"o.



\begin{thebibliography}{10}
\providecommand{\url}[1]{#1}
\csname url@samestyle\endcsname
\providecommand{\newblock}{\relax}
\providecommand{\bibinfo}[2]{#2}
\providecommand{\BIBentrySTDinterwordspacing}{\spaceskip=0pt\relax}
\providecommand{\BIBentryALTinterwordstretchfactor}{4}
\providecommand{\BIBentryALTinterwordspacing}{\spaceskip=\fontdimen2\font plus
\BIBentryALTinterwordstretchfactor\fontdimen3\font minus
  \fontdimen4\font\relax}
\providecommand{\BIBforeignlanguage}[2]{{%
\expandafter\ifx\csname l@#1\endcsname\relax
\typeout{** WARNING: IEEEtran.bst: No hyphenation pattern has been}%
\typeout{** loaded for the language `#1'. Using the pattern for}%
\typeout{** the default language instead.}%
\else
\language=\csname l@#1\endcsname
\fi
#2}}
\providecommand{\BIBdecl}{\relax}
\BIBdecl

\bibitem{Tretyakov2003a}
S.~A. Tretyakov, S.~Maslovski, and P.~A. Belov, ``An analytical model of
  metamaterials based on loaded wire dipoles,'' \emph{{IEEE} Trans. Antennas
  Propag.}, vol.~51, no.~10, pp. 2652--2658, 2003.

\bibitem{Alu2010b}
A.~Al\`u and S.~Maslovski, ``Power relations and a consistent analytical model
  for receiving wire antennas,'' \emph{{IEEE} Trans. Antennas Propag.},
  vol.~58, no.~5, pp. 1436--1448, 2010.

\bibitem{Alu2009}
A.~Al\`u and N.~Engheta, ``Cloaking a sensor,'' \emph{Phys. Rev. Lett.}, vol.
  102, p. 233901, 2009.

\bibitem{Alu2010a}
------, ``Cloaking a receiving antenna or a sensor with plasmonic
  metamaterials,'' \emph{Metamaterials}, vol.~4, pp. 153--159, 2010.

\bibitem{Green1966}
R.~B. Green, ``Scattering from conjugate-matched antennas,'' \emph{{IEEE}
  Trans. Antennas Propag.}, vol.~14, no.~1, pp. 17--21, 1966.

\bibitem{Kahn1965}
W.~Kahn and H.~Kurss, ``Minimum-scattering antennas,'' \emph{{IEEE} Trans.
  Antennas Propag.}, vol.~13, no.~5, pp. 671--675, 1965.

\bibitem{Andersen2005}
J.~B. Andersen and A.~Frandsen, ``Absorption efficiency of receiving
  antennas,'' \emph{{IEEE} Trans. Antennas Propag.}, vol.~53, no.~9, pp.
  2843--289, 2005.

\bibitem{Kwon2009}
D.-H. Kwon and D.~M. Pozar, ``Optimal characteristics of an arbitrary receive
  antenna,'' \emph{{IEEE} Trans. Antennas Propag.}, vol.~57, no.~12, pp.
  3720--3727, 2009.

\bibitem{Tretyakov1996}
S.~A. Tretyakov, F.~Mariotte, C.~R. Simovski, T.~G. Kharina, and J.~P. Heliot,
  ``Analytical antenna model for chiral scatterers: comparison with numerical
  and experimental data,'' \emph{{IEEE} Trans. Antennas Propag.}, vol.~44,
  no.~7, pp. 1006--1014, 1996.

\bibitem{Alitalo2011}
P.~Alitalo, A.~O. Karilainen, T.~Niemi, C.~R. Simovski, and S.~A. Tretyakov,
  ``Design and realisation of an electrically small {Huygens} source for
  circular polarisation,'' \emph{IET Microw. Antennas Propag.}, vol.~5, no.~7,
  pp. 783--789, May 2011.

\bibitem{Karilainen2011}
A.~O. Karilainen, P.~Alitalo, and S.~A. Tretyakov, ``Chiral antenna element as
  a low backscattering sensor,'' in \emph{Proc. 5th European Conf. Antennas
  Propag. (EUCAP)}, Rome, Italy, Apr. 11-15 2011, pp. 1983--1986.

\bibitem{antenniteoria}
I.~Lindell and K.~Nikoskinen, \emph{Antenniteoria}.\hskip 1em plus 0.5em minus
  0.4em\relax Espoo, Finland: Otatieto, 1995, in Finnish.

\bibitem{Balanis2005}
C.~A. Balanis, \emph{Antenna Theory: Analysis and Design}, 3rd~ed.\hskip 1em
  plus 0.5em minus 0.4em\relax Hoboken, NJ: Wiley, 2005.

\bibitem{Serdyukov2001}
A.~Serdyukov, I.~Semchenko, S.~Tretyakov, and A.~Sihvola,
  \emph{Electromagnetics of Bi-anisotropic Materias, Theory and
  Applications}.\hskip 1em plus 0.5em minus 0.4em\relax Amsterdam, The
  Netherlands: Gordon and Breach, 2001.

\bibitem{Lindell2009}
I.~V. Lindell, A.~Sihvola, P.~Yl\"a-Oijala, and H.~Wall\'en, ``Zero
  backscattering from self-dual objects of finite size,'' \emph{{IEEE} Trans.
  Antennas Propag.}, vol.~57, no.~9, pp. 2725--2731, 2009.

\bibitem{Lindell1996}
I.~V. Lindell, \emph{Methods for Electromagnetic Field Analysis}.\hskip 1em
  plus 0.5em minus 0.4em\relax Wiley-IEEE Press, 1996.

\bibitem{Karilainen2011c}
A.~O. Karilainen and S.~A. Tretyakov, ``Zero-backscattering self-dual object
  from two chiral particles,'' in \emph{Proc. Metamaterials 2011}, Barcelona,
  Spain, 10-13 Oct. 2011, pp. 405--407.

\bibitem{Wheeler1958}
H.~A. Wheeler, ``The spherical coil as an inductor, shield, or antenna,''
  \emph{Proc. IRE}, vol.~46, no.~9, pp. 1595--1602, 1958.

\bibitem{Thal2006}
H.~L. Thal, ``New radiation {Q} limits for spherical wire antennas,''
  \emph{{IEEE} Trans. Antennas Propag.}, vol.~54, no.~10, pp. 2757--2763, 2006.

\end{thebibliography}


\end{document}